\documentclass[aps,reprint,twocolumn,showpacs,superscriptaddress,10pt]{revtex4-1}
\usepackage{graphicx}
\usepackage{dcolumn}
\usepackage{bm}
\usepackage{amsmath}
\usepackage{epstopdf}
\usepackage{lipsum}
\usepackage{siunitx}

\newcommand{\be}[1]{\begin{equation}\label{#1}}
\newcommand{\ee}{\end{equation}}
\makeatletter
\let\Hy@backout\@gobble
\makeatother
\begin{document}

\title{Intrinsic permanent magnetic characteristics beyond
  Ce$_{2}$Co$_{17}$ with Ce/Co substitutions}
	
\author{Manish K. Kashyap}
\affiliation{Department of Physics, Kurukshetra University, Kurukshetra 136119 (Haryana), INDIA } 
\affiliation{Ames Laboratory, U.S. Department of Energy, Iowa State University, Ames, Iowa 50011-3020, USA } 
\author{Timothy A. Hackett}
\affiliation{Ames Laboratory, U.S. Department of Energy, Iowa State University, Ames, Iowa 50011-3020, USA } 
\affiliation{Department of Biochemistry, University of Nebraska, Lincoln, Nebraska 68588-0664, USA} \author{Ed Moxley}
\affiliation{Ames Laboratory, U.S. Department of Energy, Iowa State University, Ames, Iowa 50011-3020, USA } 
\author{Sarvesh Kumar}
\affiliation{Inter University Accelerator Centre, Aruna Asaf Ali Marg, New Delhi 110067, INDIA } 
\author{D. Paudyal}
\affiliation{Ames Laboratory, U.S. Department of Energy, Iowa State University, Ames, Iowa 50011-3020, USA } 
\author{B. N. Harmon}
\affiliation{Ames Laboratory, U.S. Department of Energy, Iowa State University, Ames, Iowa 50011-3020, USA } 
\affiliation{Department of Physics and Astronomy, Iowa State University, Ames, Iowa 50011-3160, USA } 


\begin{abstract} 
By understanding the small easy axis magnetocrystalline anisotropy
energy (MAE) of hexagonal Ce$_{2}$Co$_{17}$, an attempt has been made
to improve anisotropy and consequently to obtain better
characteristics for a high energy permanent magnet via site selective
substitutional doping of Ce/Co with suitable elements. The present
calculations of the electronic and magnetic properties of
Ce$_{2}$Co$_{17}$ and related substituted compounds have been
performed using the full potential linear augmented plane wave
(FPLAPW) method within the generalized gradient approximation
(GGA). Sm-substituted compounds were simulated using Coulomb corrected
GGA (GGA+$U$) to provide a better representation of energy bands due to
the strongly correlated Sm-$f$ electrons. The formation energies for all
substituted compounds are found to be negative which indicate their
structural stability. Of the substitutions, Zr substitution at the Co-dumbbell
site enhances uniaxial anisotropy of Ce$_{2}$Co$_{17}$. Furthermore,
Sm-substitution at Ce-$2c$ sites favors incremental MAE whereas a
La-substitution at both $2b$- and $2c$-sites depletes the tiny MAE in
Ce$_{2}$Co$_{17}$. These observed trends in the MAE have been examined in terms of contributions from various electronic
states. Finally, Ce$_2$Zr$_2$Co$_{15}$ and SmCeCo$_{17}$ are foreseen
as suitable materials for designing permanent magnets derived from the
crystal lattice structure of hexagonal Ce$_{2}$Co$_{17}$.
\end{abstract} 
	
	
\maketitle
	
\section{Introduction}
	
Permanent magnetic materials play an important role in improving the
efficiency, sustainability, and performance of commercial products in
electric power generation, transportation, and other energy-use
sectors of the global economy. Besides a high Curie Temperature
$(T_{c})$, desirable permanent-magnets (P-Ms) must have a high energy
product $BH_\text{max}$ and should be cost effective. $BH_\text{max}$ is
related to remnant magnetization $(M_{r})$ and coercivity $(H_{c})$
which originate from the magnetocrystalline anisotropy energy
(MAE). The MAE is one key factor for determining the easiness of a
particular magnetic material to become magnetized in one direction,
while being resistant to magnetization in other directions
\cite{Herbst, PhysRevB.29.4176}. The magnetic materials with high
$H_{c}$, uniaxial MAE, and $BH_\text{max}$ have extraordinary
applications in high-tech media and ferromagnetic electrodes of
spintronic devices with high magnetic-noise immunity and thermal
stability. Due to economic and geopolitical issues, there is an
incentive to produce P-Ms free from critical (expensive) rare earth
elements. Ce-based magnets may be a prime choice in this direction due
to their non-criticality, relatively low price, and the abundance of
Ce \cite{Herbst}.
	
Numerous experimental and theoretical efforts have been made to
enhance the MAE of Ce$_{2}$Co$_{17}$ by various
means~\cite{ke2016prbA,*ke2016a, PhysRevB.19.2704, shen1999jap, Zhi,
  fujii1982jap}. Streever \cite{PhysRevB.19.2704} used their NMR
results to evaluate the spin-orbit contribution to the magnetic
anisotropy of Co atoms in $R$Co$_{5}$ compounds. They found the easy
c-axis Co anisotropy arises from the $2c$ sites. Ke et
al.~\cite{ke2016prbA,*ke2016a} investigated the origin of MAE in doped
Ce$_{2}$Co$_{17}$ and confirmed that the dumbbell sites have a very
negative contribution to the MAE in Ce$_2$Co$_{17}$, while the MAE is
enhanced by replacing Co dumbbell sites with a pair of Fe or Mn atoms.
X-ray diffraction (XRD) measurements \cite{shen1999jap} on
magnetically aligned Ce$_{2}$Co$_{17-x}$Al$_{x}$ powders with $x=0-3$
exhibit an easy-axis of magnetic anisotropy at room
temperature. Substitution of Al for Co leads to a change of the
magnetocrystalline anisotropy of the Co sublattice from the basal
plane to c axis. The anisotropy increases with Al concentration (x),
goes through a maximum value of 17.7 kOe at $x=2$ and then
decreases. Sun et al. \cite{Zhi} prepared Ce$_{2}$Co$_{17-x}$Mn$_{x}$
($x$ = 0-4) by arc melting in argon atmosphere and found by XRD
measurements that all these magnetically aligned powered samples
exhibit an easy axis magnetic anisotropy at room temperature,
attaining a maximum value of 30.1 kOe at $x=2$.
	
Sm-Co magnets are strong permanent magnets developed in the early
1960s and mainly exist in three hard magnetic phases: SmCo$_{5}$,
SmCo$_{7}$, and Sm$_{2}$Co$_{17}$. Sm$_{2}$Co$_{17}$ can be derived
from the SmCo$_{5}$ structure and in this process, Co enrichment with
respect to SmCo$_{5}$ yields an increase in magnetization and
$BH_\text{max}$ but at a cost of the MAE \cite{Gut,Larson}. In a
similar way, the hexagonal Ce$_{2}$Co$_{17}$ system (2-17 series)
containing four Co sublattices defined by Wyckoff sites $12k$, $12j$,
$6g$, and $4f$, can also be obtained from CeCo$_{5}$ (hexagonal
CaCu$_{5}$ structure with 191:P6/mmm space group) by replacing every
third Ce atom by a pair of Co-atoms \cite{ke2016prbA,*ke2016a}. It is
clear from the example of Sm magnets, 1-5 systems have better
anisotropy but low magnetization due to a small number of Co
atoms. Therefore, it is preferable to work with 2-17 systems if
control over decreasing MAE can be achieved.

Ke et al.~\cite{ke2016prb} studied $R$(Fe$_{1-x}$Co$_x$)$_{11}$Ti$Z$
($R$ = Y and Ce; $Z$ = H, C, and N) systems for identifying improved
magnetic properties via saturation magnetization $M$, Curie
temperature $T_C$, and magnetocrystalline anisotropy energy coincident
with reduction of critical materials in permanent magnets. They found
the interstitial C doping significantly increases the uniaxial
anisotropy in Ce(Fe$_{1-x}$Co$_x$)$_{11}$Ti for $0.7 <x< 0.9$ , which
may provide the best combination of all three intrinsic magnetic
properties for permanent applications. Nd$_2$Fe$_{14}$B-based sintered
magnets were investigated extensively and used widely since being
discovered in 1984~\cite{sagawa1984jap,croat1984jap}. Tian et
al.~\cite{tian2018aa} recently discovered the practical value of
$R_2$Fe$_{14}$B ($R$ = Pr, Nd) magnets can be improved by proper
doping of nitrogen into some interstitial sites. Ke and van
Schilfgaarde~\cite{ke2015prb} explained when Li is substituted by Fe
in lithium nitride, the resultant compound; Li$_2$(Li$_{1-x}$Fe$_x$)N
behaves in many aspects similar to that of rare earth element. In
addition, they unveiled an analytical model to describe the
magnetocrystalline anisotropy energy (MAE) in solids as a function of
band filling. The MAE decomposed into a sum of transitions between
occupied and unoccupied pairs. This model also yielded the MAE in good
qualitative agreement with its value from first-principles
calculations for Li$_2$(Li$_{1-x}T_x$)N, with $T$ =Mn, Fe, Co, and Ni
systems. These examples indicate doping substitutional replacement of
non-critical atoms is pivotal for enhancing the MAE in particular
compounds.

Pure Ce$_{2}$Co$_{17}$ exhibits poor anisotropy characteristics and has
a theoretical energy product of only $~$ 31 MGOe \cite{book},
therefore it is not a suitable P-M.  Sun et al.~\cite{sun2017jsnm}
also established poor performance of CeCo$_5$-based compounds;
Ce(Co$_{0.73}$Cu$_{0.135}$Fe$_{0.135}$)$_{5.35}$ and
Ce(Co$_{0.73}$Cu$_{0.135}$Fe$_{0.135}$)$_{5.55}$ for permanent magnets
due to the formation of Ce$_2$Co$_{17}$ phase with
$T_\text{C}<\SI{20}{\celsius}$.  Improvement in the anisotropy of
Ce$_{2}$Co$_{17}$ is expected by suitable substitution of Co as observed
in the literature, perhaps to a point beyond which it is auspicious
for PM fabrication. Due to the presence of inequivalent sites of the
Co and Ce atoms in Ce$_{2}$Co$_{17}$, it may be possible to substitute
these atoms by foreign atoms and design new compounds with better
magnetic properties and enhanced MAE. The main aim of this study is to
theoretically investigate the change in the MAE via this type of
substitution. For this purpose, the full potential density functional
approach has been selected for studying Co substitution by Zr/Ti, and
Ce substitution by Sm/La.

\section{Computational Approach}
	
The base compound; Ce$_{2}$Co$_{17}$ crystallizes in the
Th$_{2}$Ni$_{17}$-type (space group: 194; P63/mmc) structure
\cite{Khan}. For spin polarized calculations of pristine
Ce$_{2}$Co$_{17}$ and its $4f$ site substitution, we considered the
lattice parameters of Ce$_{2}$Co$_{17}$ \cite{Gut} whereas for
$2b$/$2c$ substitution, we have modeled the lattice parameters as an
average of lattice parameters of Ce$_{2}$Co$_{17}$ and
Sm$_{2}$Co$_{17}$\cite{BUSCHOW1968323}/La$_{2}$Co$_{17}$
\cite{Chang2003}. The calculations were performed using density
functional theory (DFT) \cite{PhysRevB.26.4571} based upon the
full-potential linearized augmented plane wave (FPLAPW) method
\cite{PhysRevB.64.195134} as implemented in the WIEN2k crystal program
\cite{wien2k}. The generalized gradient approximation (GGA) under the
parameterization of Perdew-Burke-Ernzehof (PBE)
\cite{PhysRevLett.77.3865} was used to construct the
exchange-correlation (XC) functionals for all compounds except
Sm-related ones. For Sm substitutions, XC functional were considered
within GGA+$U$ to correct on-site f-electron correlation by taking
$U_{eff}$ = $U - J$ = 6.0 eV \cite{Vlad}, which includes the Coulomb
parameter, $U=6.7$ eV and exchange parameter $J= 0.7$ eV for the Sm-f
electrons. We used $R_{mt}k_{max}$ = 8.5 for basis functions, and the
cut-off energy and charge convergence criteria were set to
$~10^{-6}$Ry and $~10^{-6}$e, respectively, due to the sophisticated
MAE calculations involved. The k-space integration was performed using
the modified tetrahedron method \cite{PhysRevB.49.16223}.
Self-consistency was obtained using 168 k-points in the irreducible
Brillouin zone (IBZ). A scalar relativistic calculation, with GGA or
GGA+$U$ formalisms, was first performed to obtain a self-consistent
potential. After fixing this potential, relativistic effects were
included with the second variational treatment of spin-orbit
coupling~\cite{koelling1977jpcs}. The MAE stems from the orbital
contribution to the magnetic moment mainly. We calculated the MAE by
using the magnetic force theorem \cite{book1}. The MAE constant ($K$),
which is MAE/volume of the cell, is then calculated from the
expression:

\begin{equation}
  K= \sum_{j,k}^{occu}\epsilon_{j}(\hat{n}_{2},k) - \sum_{j,k}^{occu}\epsilon_{j}(\hat{n}_{1},k)
  \label{eq1}
\end{equation}

Here $\epsilon_{j}$ is the Kohn-Sham eigenvalue evaluated for each
magnetization orientation.  In Eq. \ref{eq1}, $j$ labels occupied
states and the $k$ is a particular $k$-point in the Brillouin zone, $\hat{n}_{1}$ is $[001]$, i.e., $c$-axis and $\hat{n}_{2}$ is taken along the $[100]$ direction. Here, the positive/negative value of E
indicates the uniaxial/easy-plane anisotropy. The calculations of MAE
demand a fine energy resolution and dense mesh of in k-space, hence,
we set up a $17\times17\times15$ k-point mesh for all the anisotropy
calculations in the spin orbit coupling environment.
	
\section{Results and Discussion}
	
\begin{figure}[!h]
  \centerline{\includegraphics[width=1.0\linewidth,angle=0]{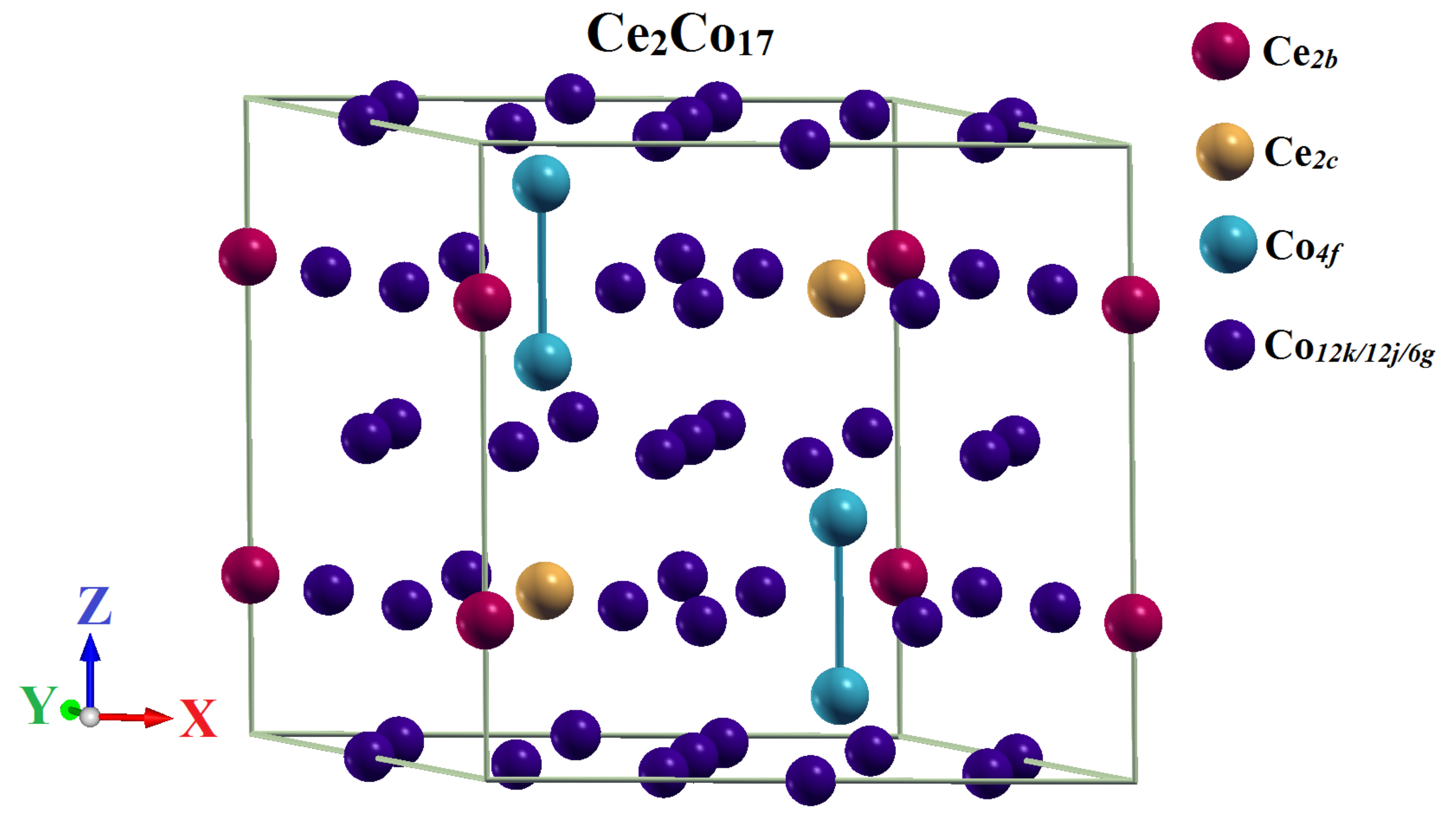}}
  \caption{\label{fig1} Schematics for Ce$_{2}$Co$_{17}$ showing Co
    dumbbell at $4f$ sites and Co $2c$/$2b$ sites, available for
    substitution.  }
\end{figure}

\begin{figure}[!t]
  \centerline{\includegraphics[width=1.0\linewidth]{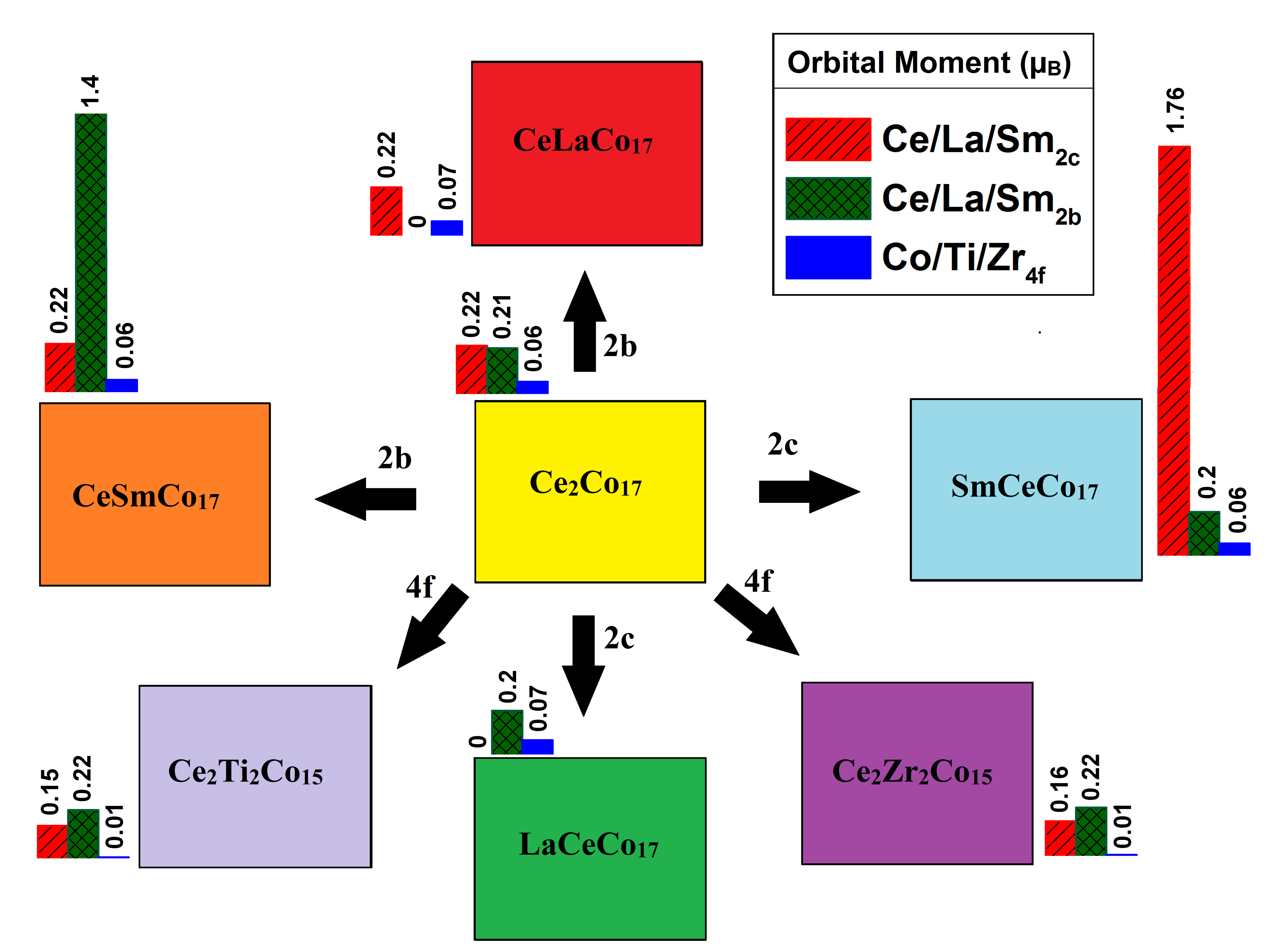}}
  \caption{\label{fig2} Symbolic site specification and nomenclature
    for Ce and Co substitution in Ce$_{2}$Co$_{17}$ at $4f$, $2b$ and $2c$. }
\end{figure}

The identification of effective sites for substitution is needed to
enhance the MAE of Ce$_{2}$Co$_{17}$. The literature reveals that
previous researchers substituted Co$_{4f}$ (dumbbell sites) as shown
in Fig.\ref{fig1} by foreign elements and observed a significant
change in the MAE \cite{ke2016prbA, shen1999jap, Zhi,
  fujii1982jap}. Along the same line, we started with similar
substitution by Ti and Zr.  Further, Ce atoms in Ce$_{2}$Co$_{17}$,
presented at $2b$ and $2c$ sites, as highlighted by different colors
in Fig. \ref{fig1}, are also expected to contribute significantly,
thus we also identified these sites and investigated the substitutions
at these by Sm and La. The symbolic representation used for resultant
alloys obtained by 4$f$, $2b$ and $2c$ site substitutions is depicted
in Fig. \ref{fig2}. The orbital moments as obtained by proper
substitutions are also mentioned to show the importance of the alloy
within this figure exclusively.
	
\begin{figure}[!h]
  \centerline{\includegraphics[width=1.0\linewidth]{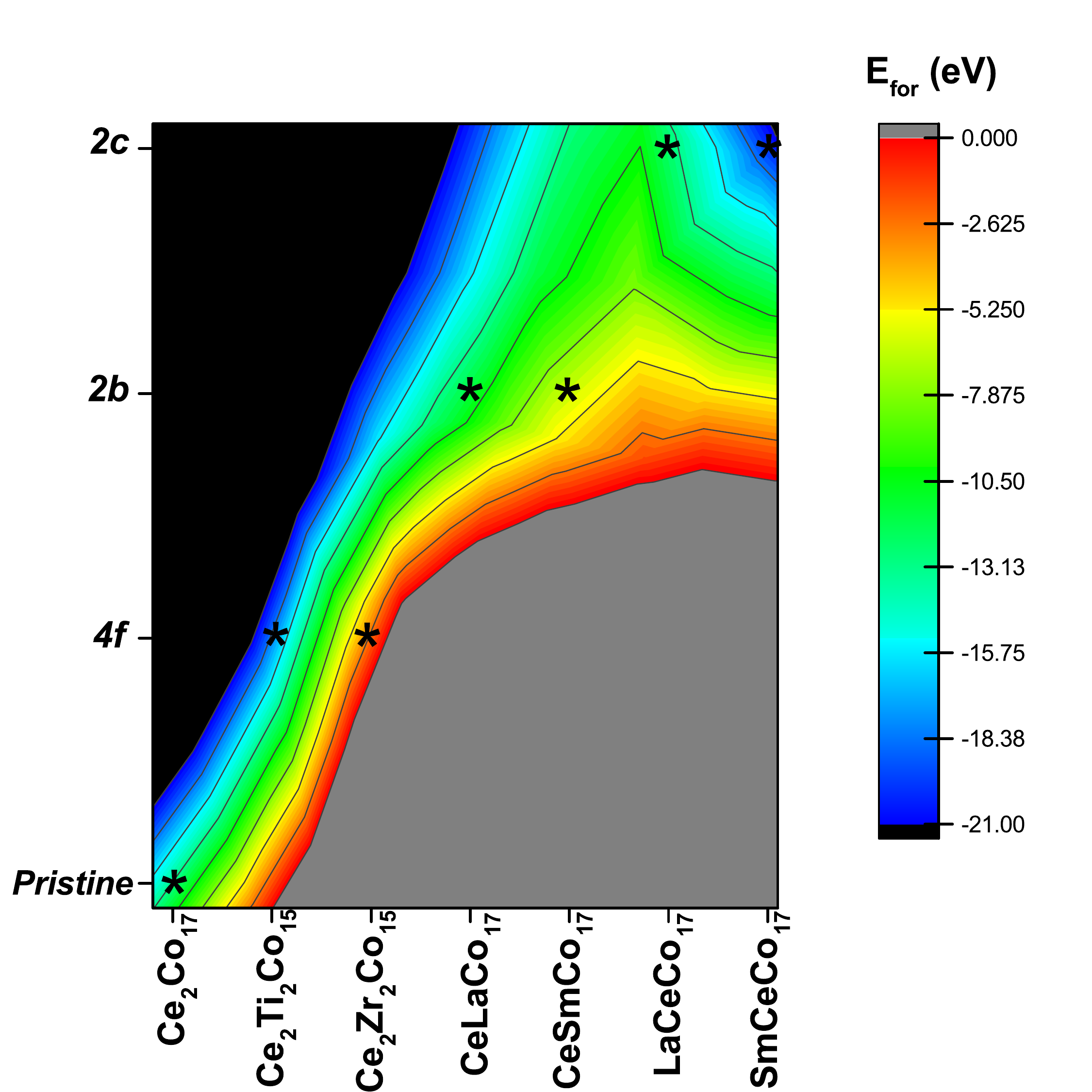}}
  \caption{\label{fig3} The contour plot for $E_{for}$ of the pristine
    Ce$_{2}$Co$_{17}$ and $4f$/$2b$/$2c$ substituted alloys (Here $*$
    represents the actual magnitude of $E_{for}$ with a particular
    substitution).}
\end{figure}

In order to estimate the stability of the computed structure with
respect to Ce$_{2}$Co$_{17}$, the formation energies; $E_{for}$(defined
by the difference of the accumulated sum of equilibrium energies of
all constituent atoms from the equilibrium energy of the resultant
compound) was estimated which are depicted in the contour plot
(Fig. \ref{fig3}). We observe that out of the studied compounds, only
Ce$_{2}$Ti$_{2}$Co$_{15}$ and SmCeCo$_{17}$ are more stable and easy to
synthesize as compared to Ce$_{2}$Co$_{17}$.  However, all compounds are
stable as verified from the negative values of formation energies
which is a good indication of the feasibility of these compounds
experimentally as well.
	
\begin{figure*}[!t]
  \centerline{\includegraphics[width=0.95\linewidth]{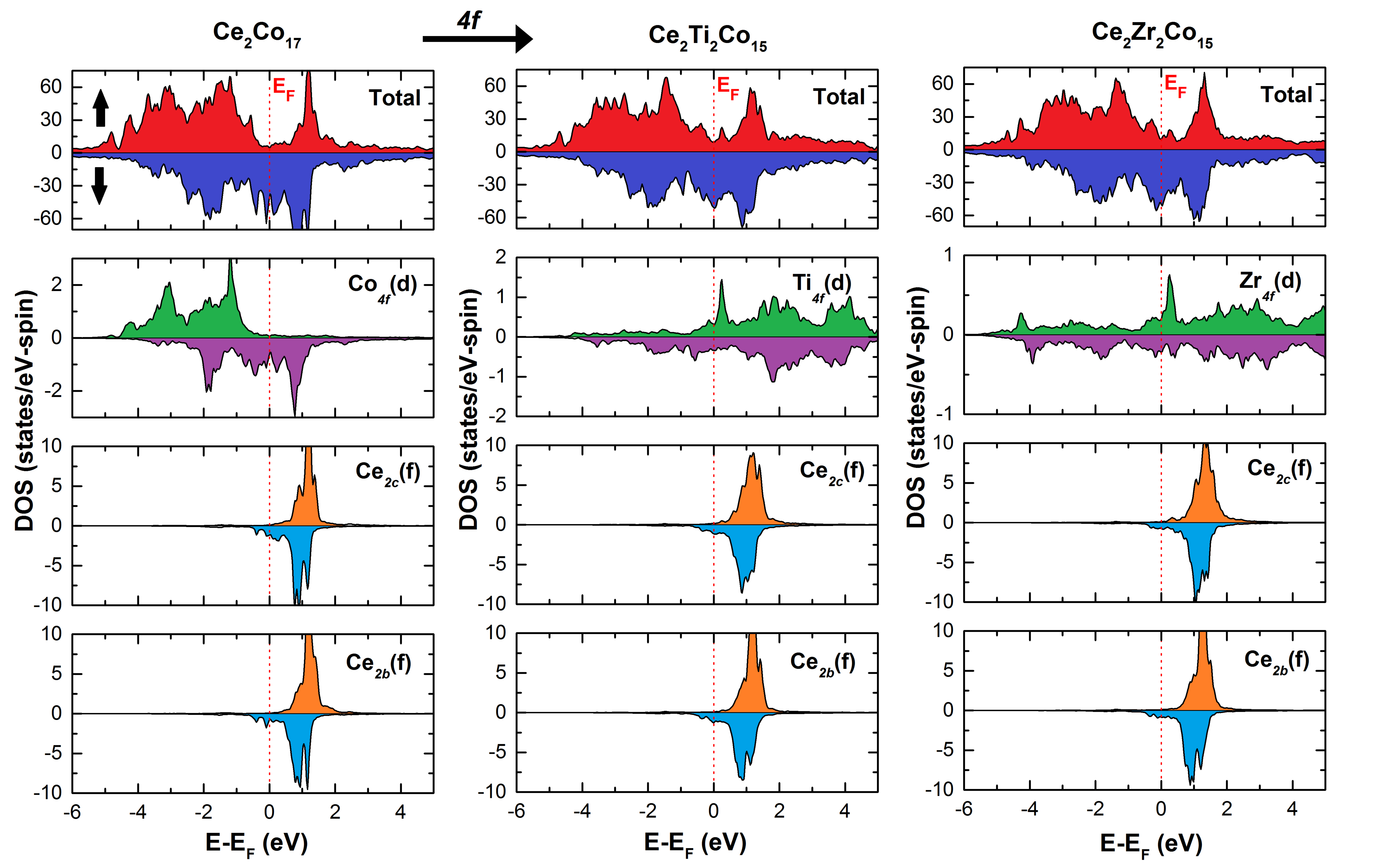}}
  \caption{\label{fig4} Total and atom resolved DOS at $2b$, $2c$ and
    $4f$ sites in Ce$_{2}$Co$_{17}$, Ce$_{2}$Ti$_{2}$Co$_{15}$ and
    Ce$_{2}$Zr$_{2}$Co$_{15}$ }
\end{figure*}

\begin{figure}[!bht]
  \centerline{\includegraphics[width=0.75\linewidth]{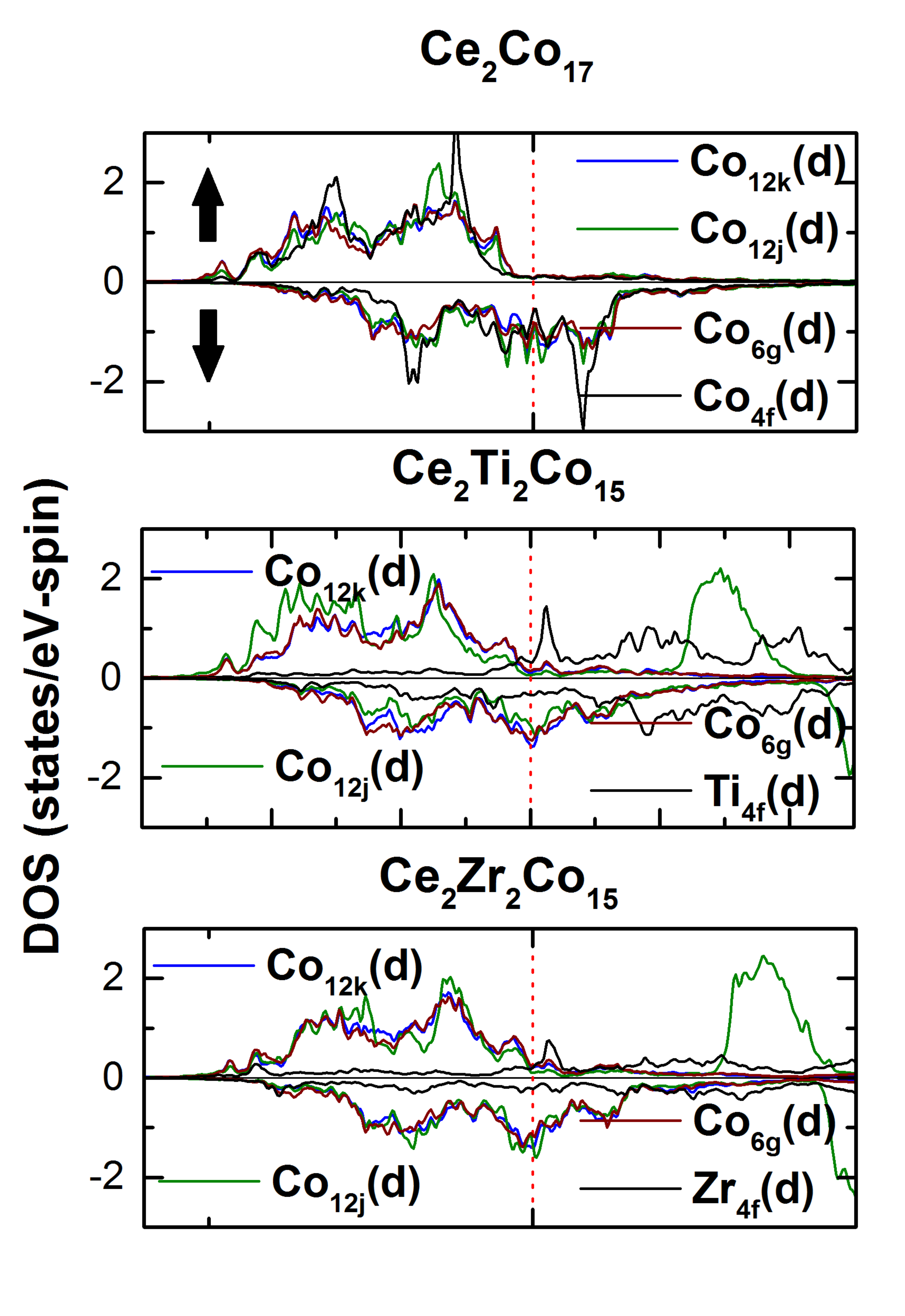}}
  \caption{\label{fig5} Partial DOS at all types of Co-sites in
    Ce$_2$Co$_{17}$, Ce$_2$Ti$_2$Co$_{15}$, and Ce$_2$ZrCo$_{15}$.}
\end{figure}

The total density of states (DOS) and atom resolved density of states
at $4f$, $2b$ and $2c$ sites in Ce$_{2}$Co$_{17}$,
Ce$_{2}$Ti$_{2}$Co$_{15}$ and Ce$_{2}$Zr$_{2}$Co$_{15}$ are depicted
in Fig.\ref{fig4}. All three total DOS are highly spin polarized,
indicating the strong magnetism in the alloys. With Ti/Zr
substitution, the states for majority spin above the Fermi level
($E_{F}$) increases due to fewer d-electrons i.e. in $3d^{2}$/$4d^{2}$
configurations as compared to Co with $3d^{7}$ configuration.  As
expected, Ce-$f$ states contribute mainly in bands above $E_{F}$. The
states in the vicinity of $E_{F}$ are required for electrical
conductivity. The Co-$d$ states at various sites for Ti/Zr
substitution in Ce$_{2}$Co$_{17}$ are analyzed in Fig.\ref{fig5}. It
is evident that $d$-DOS of Co at the 12k and 6g sites is not much
affected by Ti/Zr substitution. A drastic change in d-DOS of Co at 12j
site can be identified above $E_{F}$. However, these states are
located at energy greater than 2~eV above $E_{F}$, so these are
insignificant in the vicinity of $E_{F}$ for easy excitations. Thus
any change in total DOS is mainly due to the presence of Ti/Zr-d
states in the substituted alloys. To emphasize the significant changes
in total DOS, the atom resolved DOS at the substituted Co-site is
plotted with separate colors.
	
\begin{figure*}[!t]
  \centerline{\includegraphics[width=1.0\linewidth]{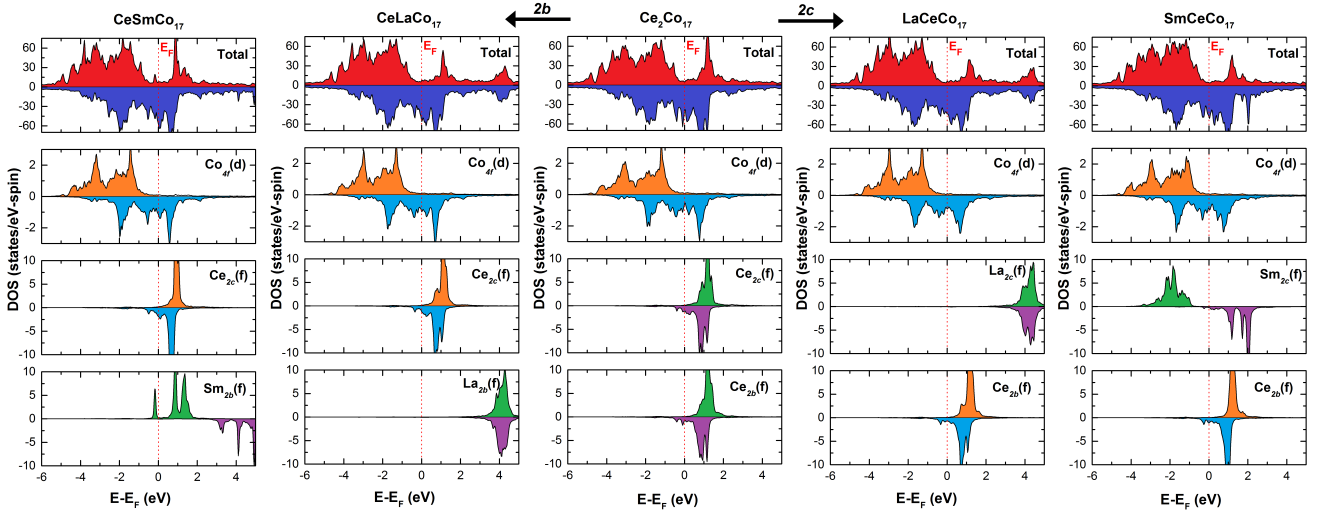}}
  \caption{\label{fig6} Total and atom resolved DOS at $2b$, $2c$ and
    $4f$ sites in CeLaCo$_{17}$, CeSmCo$_{17}$, LaCeCo$_{17}$ and
    SmCeCo$_{17}$ }
\end{figure*}

After identifying the effect of substitution of Co-atoms (at dumbbell
sites) on the electronic properties of Ce$_{2}$Co$_{17}$, we have
extended this study to Ce-site substitutions by Sm and La
atoms. Fig.\ref{fig6} depicts DOS of Ce/Sm/La-f states at $2b$ and/or $2c$
sites. On analysis, it is found that for the DOS at the dumbbell site;
Co$_{4f}$-d remains unaffected with these substitutions. Further for
substitution at $2b$ or $2c$ by La, the unoccupied La-f are unavailable in
the vicinity of $E_{F}$ which reduces the spin polarization in the
resultant system. Significant change in total DOS is governed by Sm
substitution at $2b$ and $2c$ sites. The two Sm-sites substitutions behave
quite differently. While $2c$ substitution gives rise to more occupied
Sm-f states, the $2b$-substitution produces very few Sm-f DOS in the
occupied region. The remaining Ce-atom in the unit cell on the other
site, has similar type of DOS for both $2b$/$2c$ substitution. Further,
these are found to be little perturbed compared to the case of
pristine Ce$_{2}$Co$_{17}$. To confirm any change on Co-sites after
substitution, we also analyzed all the Co sites (not shown for
brevity). But surprisingly, all states are found to possess similar
DOS as that of pristine case and do not have significant influence in
deciding the qualitative features of total DOS. However, the strong
spin polarization at each site results in better exchange spitting
in all resultant alloys.
	
\begin{figure}[!h]
  \centerline{\includegraphics[width=1.0\linewidth]{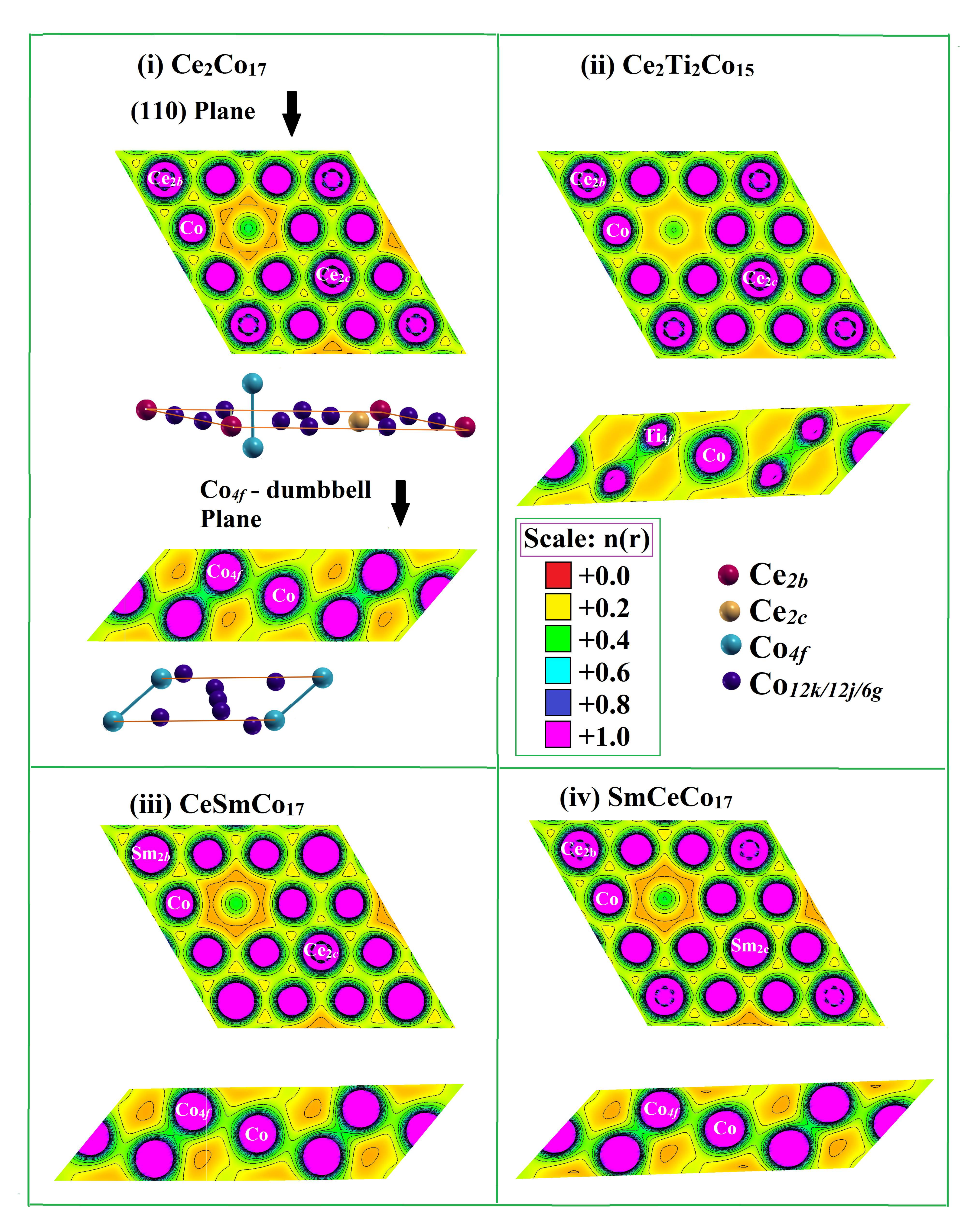}}
  \caption{\label{fig7} Total valence electronic charge density
    difference, n(r) in units of $e/\AA^{3}$ in Ce$_{2}$Co$_{17}$,
    Ce$_{2}$Ti$_{2}$Co$_{15}$, CeSmCo$_{17}$ and SmCeCo$_{17}$.  }
\end{figure}

Total valence electronic charge densities, n(r) in the (110) and
Co$_{4f}$ - dumbbell planes for (i) pristine Ce$_{2}$Co$_{17}$ (ii)
Ce$_{2}$Ti$_{2}$Co$_{15}$ (iii) CeSmCo$_{17}$ and (iv) SmCeCo$_{17}$
compounds were analyzed using the xcrysden program
\cite{Kokalj2003155} to examine the effect on the neighboring atoms
after substitution at $4f$, $2b$ and $2c$ sites in
Ce$_{2}$Co$_{17}$. Focusing on pristine Ce$_{2}$Co$_{17}$
(Fig.\ref{fig7}) the effect of symmetrical Co$_{4f}$ atoms aligned in
the dumbbell form is clearly visible, with maximum charge density
along the dumbbell axis (see green color inside hexagonal contour in
Fig.\ref{fig7} as a projection of charge density from two Co-atoms
aligned in the dumbbell form on both sides of the $(110)$ plane in
perpendicular direction). Further charge density contours in the
Co$_{4f}$ dumbbell plane for Ce$_{2}$Co$_{17}$ indicate that two Co
atoms at $4f$ sites are tightly bonded with each other. But the
substitution of Ti in place of Co atoms at $4f$ sites decreases the
charge densities along the Co$_{4f}$ dumbbell axis due to the presence
of a lesser number of d-electrons. The dumbbell Ti-atoms are not
tightly bonded and are expected to decrease the magnetism in the
resultant compound.

\begin{figure}[!h]
  \centerline{\includegraphics[width=1.0\linewidth]{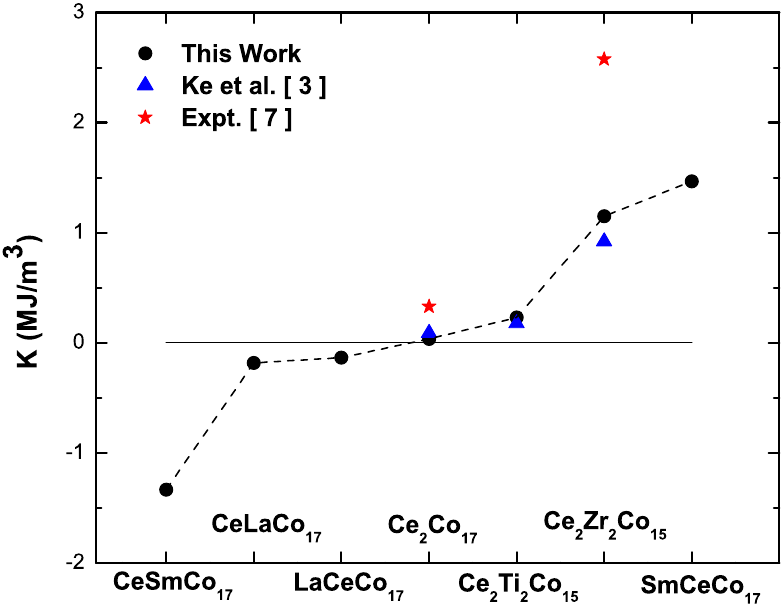}}
  \caption{\label{fig8} Calculated MAE (MJ/$m^3$) for Ce$_{2}$Co$_{17}$
    and all other substituted alloys. The corresponding experimental
    \cite{fujii1982jap} and predicted \cite{ke2016prbA,*ke2016a}
    values are also presented for comparison.  }
\end{figure}

The charge densities along the Co$_{4f}$ – dumbbell axis for
Sm-$2b$/$2c$ substitution are again enhanced. However, in these
substitutions, the effect on charge densities along Co$_{4f}$ –
dumbbell plane is modest and dumbbell Co-atoms are again tightly
bonded (as in case of Ce$_{2}$Co$_{17}$). The substituted Sm atom
behaves differently in the resultant compound which is clearly visible
from the shape of Ce and Sm atoms as depicted in Fig.\ref{fig7}(iii)
and (iv), respectively. The Ce atom tends to polarize the neighboring
Co-atoms but the Sm atom preserves its spherical shape. The two
factors, strong bonding between dumbbell Co-atoms and spherical shape
of Sm atoms seem to increase the overall magnetism. These density
plots demonstrate the importance of Co-dumbbell sites in the present
compounds not only for $4f$ but also for $2b$/$2c$ substitutions.

\begin{table*}[!t]
  \caption{Calculated total orbital ($\mu_l$), spin ($\mu_s$) and
    total moments ($\mu_t$) in $\mu_{B}$ for Ce$_{2}$Co$_{17}$ and all
    related substituted compounds} \centering
  \label{table}
  \begin{tabular}{|p{0.8in}|p{0.5in}|p{0.3in}|p{0.5in}|p{0.5in}|p{0.6in}|p{0.5in}|p{0.5in}|p{0.5in}|p{0.5in}|p{0.5in}|} \hline 
    \textbf{\newline Material} & \textbf{\newline } & \multicolumn{9}{|p{3.3in}|}{\textbf{\newline Magnetic moment(}$\mu$)} \\ \hline 
    \textbf{} & \textbf{} & \textbf{} & \textbf{\newline Ce($2c$)} & \textbf{\newline Ce($2b$)} & \textbf{\newline Co($12k$)} & \textbf{\newline Co($12j$)} & \textbf{\newline Co($6g$)} & \textbf{\newline Co($4f$)} & \textbf{\newline Inter.} & \textbf{\newline Total} \\ \hline 
    Ce${}_{2}$Co${}_{17}$ & This work\newline \newline Ke et al. \cite{ke2016prbA}\newline \newline Expt \cite{Hu199983} \newline  & $\mu$${}_{s}$ \newline $\mu$${}_{l}$ \newline $\mu$${}_{t}$ \newline $\mu$${}_{s}$ \newline $\mu$${}_{l}$ \newline $\mu$${}_{t}$ \newline $\mu$${}_{t}$ & -0.58\newline 0.22\newline -\newline -0.84\newline 0.38\newline -\newline - & -0.63\newline 0.21\newline -\newline -0.90\newline 0.42\newline -\newline - & 1.60\newline 0.08\newline -\newline 1.51\newline 0.10\newline -\newline -\newline  & 1.78\newline 0.09\newline -\newline 1.56\newline 0.11\newline -\newline -\newline  & 1.55\newline 0.07\newline -\newline 1.51\newline 0.08\newline -\newline -\newline  & 1.74\newline 0.06\newline -\newline 1.65\newline 0.07\newline -\newline - & -4.04\newline -\newline -\newline -1.1 \newline -\newline -\newline - & 23.58\newline 1.78\newline 25.36\newline 23.40\newline 2.43\newline 25.8\newline 25.4 \\ \hline 
    Ce${}_{2}$Zr${}_{2}$Co${}_{15}$\newline Zr ($4f$) & This work\newline  & $\mu$${}_{s}$ \newline $\mu$${}_{l}$\newline $\mu$${}_{t}$\newline  & -0.47\newline 0.20\newline - & -0.52\newline 0.47\newline - & 1.22\newline 0.07\newline - & 1.34\newline 0.10\newline - & 1.07\newline 0.08\newline - & -0.19\newline 0.01\newline - & -3.36\newline -\newline - & 15.52\newline 1.95\newline 16.37 \\ \hline 
    Ce${}_{2}$Ti${}_{2}$Co${}_{15}$\newline Ti ($4f$) & This work\newline  & $\mu$${}_{s}$ \newline $\mu$${}_{l}$\newline $\mu$${}_{t}$\newline  & -0.58\newline 0.15\newline - & -0.66\newline 0.22\newline - & 1.27\newline 0.07\newline - & 1.92\newline 0.10\newline - & 1.16\newline 0.07\newline - & -0.52\newline 0.01\newline - & -4.22\newline -\newline - & 16.07\newline 1.62\newline 17.69 \\ \hline 
    CeLaCo${}_{17}$\newline La ($2b$) & This work\newline  & $\mu$${}_{s}$ \newline $\mu$${}_{l}$\newline $\mu$${}_{t}$\newline  & -0.62\newline 0.22\newline - & -0.12\newline 0.00\newline - & 1.67\newline 0.09\newline - & 1.56\newline 0.12\newline - & 1.63\newline 0.08\newline - & 1.81\newline 0.07\newline - & -4.12\newline -\newline - & 25.18\newline 1.86\newline 27.04 \\ \hline 
    CeSmCo${}_{17}$\newline Sm ($2b$) & This work\newline  & $\mu$${}_{s}$ \newline $\mu$${}_{l}$\newline $\mu$${}_{t}$\newline  & -0.60\newline 0.22\newline -0.48 & 4.82\newline -1.40\newline 3.42 & 1.65\newline 0.09\newline 1.74 & 1.55\newline 0.11\newline 1.66 & 1.60\newline 0.08\newline 1.68 & 1.77\newline 0.06\newline 1.83 & -3.44\newline -\newline -3.44 & 30.08\newline 0.38\newline 30.46 \\ \hline 
    LaCeCo${}_{17}$\newline La ($2c$) & This work\newline  & $\mu$${}_{s}$ \newline $\mu$${}_{l}$\newline $\mu$${}_{t}$\newline  & -0.11\newline 0.00\newline - & -0.67\newline 0.20\newline - & 1.67\newline 0.08\newline - & 1.56\newline 0.10\newline - & 1.63\newline 0.08\newline - & 1.81\newline 0.07\newline - & -4.18\newline -\newline - & 25.09\newline 1.66\newline 26.75 \\ \hline 
    SmCeCo${}_{17}$\newline Sm ($2c$) & This work\newline  & $\mu$${}_{s}$ \newline $\mu$${}_{l}$\newline $\mu$${}_{t}$\newline  & 5.02\newline -1.76\newline 3.26 & -0.66\newline 0.20\newline - & 1.66\newline 0.08\newline - & 1.56\newline 0.11\newline - & 1.63\newline 0.08\newline - & 1.79\newline 0.06\newline - & -3.32\newline -\newline - & 30.56\newline -0.06\newline 30.50 \\ \hline 
  \end{tabular}
\end{table*}

The atom resolved spin and orbital moments at each site are listed in
Table-I. It is clear that $4f$ site substitution decreases the total
moment of Ce$_{2}$Co$_{17}$ whereas $2b$/$2c$ substitution increases the
same. The magnetization is found to be in excellent agreement with the
previous experiment~\cite{Hu199983} and
calculation~\cite{ke2016prbA,*ke2016a}. The Co spin moments on each
site are similar $\approx$ 1.5-1.8 $\mu_{B}$. However on $4f$
substitution, the Ti/Zr moment aligns antiparallel to Co moment at
other sites which causes a significant decrease in total spin moment
of the resultant alloy.

Regarding $2b$/$2c$ substitution by La, the La orbital moment not only
drops down to zero but also its spin moment gets reduced. There is not
much change either in orbital or in spin moment observed for Ce as
well as Co atoms. However, this La-substitution slightly increases the
total moment due to dilution of antiparallel alignment of La and Co
spin moments. On the other hand, a significant change is observed for
Ce substitution by Sm in Ce$_{2}$Co$_{17}$. The Sm atom with $4f^{5}$
configuration has the ability to enhance the total moment. The large
increase in total magnetic moment is the consequence of the enhanced
net Sm moment of 3.42/3.26 $\mu_{B}$ at $2b$/$2c$ sites in
CeSmCo$_{17}$/SmCeCo$_{17}$. Furthermore, the Sm moment aligns
antiparallel/parallel to the Ce/Co spin moment which also favors the
increase in total moment. Focusing on orbital moment only, we have
observed that all atoms have very small orbital moments except for
Sm-atoms in CeSmCo$_{17}$ and SmCeCo$_{17}$ which is expected due to
Sm-$4f^{5}$ states.

Finally, to establish the potential of the studied alloys in permanent
magnets, the MAE ($K$) for the studied systems were examined in
Fig.\ref{fig8}. The value of $K$ for Ce$_{2}$Co$_{17}$ is poor
($\approx$ 0.03 MJ/$m^3$) as confirmed from our calculations and there
is no scope to fabricate permanent magnets based on pristine
Ce$_{2}$Co$_{17}$ alone. Further the previous calculations by Ke et
al. ~\cite{ke2016prbA,*ke2016a} and experimental work by Hu et
al. ~\cite{Hu199983} also reported a tiny value (0.09 MJ/$m^3$ and
0.32 MJ/$m^3$, respectively) of the same. The $4f$- site substitution
by Zr increases the MAE of Ce$_{2}$Co$_{17}$ significantly. We found
the MAE for this substitution (Ce$_{2}$Zr$_{2}$Co$_{15}$) as 1.15
MJ/$m^3$ which is within reasonable agreement with the corresponding
experimental value of ~ 2.57 MJ/$m^3$ ~\cite{fujii1982jap} and close
to that predicted by Ke et. al. ~\cite{ke2016prbA,*ke2016a}. The
substituted Zr-atom contributes positively in the value of MAE which
strains alignment of the magnetization from all atoms along any other
magnetic field direction except for the easy c-axis direction. In one
experiment with a similar type of substitution, Al atoms in
Ce$_{2}$Al$_{2}$Co$_{15}$~\cite{shen1999jap} were found to prefer the
dumbbell sites and are responsible in increasing uniaxial
anisotropy---which also agrees with our findings. On the other hand,
the increase in MAE via similar substitution by Ti is not as effective
and leads to an increase in $K$ to 0.23 MJ/$m^3$. The difference in
MAE obtained is a consequence of the atomic size of the substituted
atom (Zr/Ti). Whereas Zr with a larger radius at dumbbell sites leads
to strong hybridization; Ti-atom with smaller size than Co results in
an ineffective coupling.

The site substitution at $2b$/$2c$ sites by La has an opposite effect
on the MAE. Rather than increasing the uniaxial anisotropy,
La-substitution brings out in-plane anisotropy of the material due to
the unavailability of La-$4f$ electrons.  In contrast to La, site
selective Sm substitution which is rich in f-electrons, can have a
considerable effect on the overall MAE of the resultant compound,
depending upon the crystal environment in the vicinity of
substitution. In $2b$ substitution, Sm atoms are found away from
Co-dumbbell sites and have impact along $xy$ plane, therefore, this
substitution also brings down the anisotropy of Ce$_{2}$Co$_{17}$ in
planner mode. However, for $2c$ substitution, Sm atoms align coaxially
with the Co-dumbbells and thus the spin alignment along the c-axis
(which is also an easy axis) is favourable. As a result, the
involvement of $4f^5$ states gives rise to increased magnetic
anisotropy in this case and the final value of $K$ reaches to 1.4
MJ/$m^3$.

\section{Conclusion}
	
In the regime of density functional theory, pristine Ce$_{2}$Co$_{17}$
and compounds with site substitution of Ce/Co atom by La,Sm/Ti,Zr have
been investigated. The Co-dumbbell sites are very crucial for the
overall MAE in the resultant compound. The calculated MAE is found
meaningful for Zr/Sm substitution at the $4f$/$2c$ site. The
enhancement of MAE in these calculations for Zr-substitution is in
accordance with the available experimental data. The high convergence
was cross-checked so that these results can be used as a benchmark in
future experiments/calculations. Sm substitution at the $2c$ site is
vital on the basis of magnetization as well as MAE obtained.  Further
experimental analysis is required to establish the candidature of
Ce$_{2}$Zr$_{2}$Co$_{15}$ and SmCeCo$_{17}$ as permanent
magnets. Keeping in mind the cost of Sm, the present analysis is
important as most of the prominent permanent magnets are made from
SmCo$_{5}$ and Sm$_{2}$Co$_{17}$. Therefore reducing significant
content of Sm by substituting cheaper Ce in the material and
preserving a large magnetic moment may open up new avenues for the
development of permanent magnets at a significant reduced cost.

\section{Acknowledgments}
	
M.K.K. would like to acknowledge UGC, New Delhi for providing Raman
Post-Doctoral Fellowship vide grant No. 5-147/2016 (IC). This research
is supported by the Critical Materials Institute, an Energy Innovation
Hub funded by the U.S. Department of Energy (US-DoE), Office of Energy
Efficiency and Renewable Energy, Advanced Manufacturing Office. The
Ames Laboratory is operated for the US-DoE by Iowa State University of
Science and Technology under Contract No. DE-AC02-07CH11358. T.A.H. is
grateful to the US-DoE for the assistantship and opportunity to
participate in the SULI program for undergraduate research.

\bibliographystyle{apsrev4-1}
\bibliography{mybibfile}
	
\end{document}